# QUASICRYSTALLINE AND CRYSTALLINE RARE-GAS CLUSTERS PRODUCED IN SUPERSONIC JETS: IMPACT OF THE JET CLUSTERING LEVEL ON CATHODOLUMINESCENCE SPECTRA

V.L. Vakula, O.G. Danylchenko, Yu.S. Doronin, G.V. Kamarchuk,

O.P. Konotop, V.N. Samovarov, A.A. Tkachenko

*B. Verkin Institute for Low-Temperature Physics and Engineering*

*of the National Academy of Sciences of Ukraine*

*47 Nauky Ave., Kharkiv, 61103, Ukraine*

E-mail: vakula@ilt.kharkov.ua

**Abstract**

The paper proposes a new approach to studying cathodoluminescence spectra of substrate-free rare-gas clusters produced in supersonic jets exhausting into a vacuum. The approach, which takes into account the fraction of the clustered substance in the jet, is applied to quantitatively analyze integrated intensities of the luminescence bands of the neutral and charged excimer complexes $(Rg_2)^*$ and $(Rg_4^+)^*$ measured for nanoclusters of three rare gases (Rg = Ar, Kr, and Xe) with average sizes ranging from 100 to 18000 atoms per clusters (diameters varying from 2 to 13 nm). The amount of the clustered substance, which affects the absolute values of integrated intensity of the bands, is shown to be proportional to the logarithm of the average size of clusters in the jet. Analysis of normalized intensities allowed us to spectroscopically find two ranges of average sizes of Ar, Kr, and Xe nanoclusters which, in accordance with the electron diffraction studies, can be assigned to quasicrystalline icosahedral and crystalline fcc structures in clusters, as well as to find the cluster size range in which both structures coexist. We show that in fcc clusters the luminescence of the neutral molecules $(Rg_2)^*$ comes from within the volume of the cluster, while the charged excimer complexes $(Rg_4^+)^*$ emit mostly from subsurface layers.

# 1. Introduction

Clusters, i.e. small aggregations of atoms or molecules which can be thought of as an intermediate link between single particles and macroscopic solid objects, are of great interest from the viewpoint of both fundamental studies and practical applications making use of new cluster technologies [1-3]. Being, in a way, *nuclei* of the macroscopic phase, they allow one to study some of the fundamental properties of bulk solids and simulate the kinetics of the processes which result in the formation of various macroscopic structural states (see, e.g., [4]). At the same time, clusters have a number of unique properties which have never been observed in bulk samples. For example, cluster properties depend greatly on their size: by varying it, it is possible to change the cluster structure and study the structural transformations [5]. Along with the structural phases typical of bulk samples, clusters may display some specific structural states, for instance, quasicrystalline structures with a 5-fold axis of symmetry, which are not reported for bulk objects [6].

Weakly bound van-der-Waals rare-gas clusters are good model objects for studying the electronic subsystem. These clusters are known to have a comparatively simple electronic structure, which, as was shown in a number of studies [7-11], is sensitive to their spatial organization. One of the most efficient methods of the cluster production is adiabatic exhaustion of a gaseous jet through a supersonic nozzle into a vacuum (see, e.g., [12]). An important advantage of this technique is the possibility of generating substrate-free clusters with physical and chemical properties not affected by a substrate or a matrix. A convenient way to study the electronic subsystem of such clusters is to employ spectroscopic methods, including the cathodoluminescence technique used in the present study. The technique allows one to effectively investigate the processes of formation of electronic excitations in clusters and their subsequent relaxation.

One of the main problems in studying clusters produced in supersonic jets is the complex composition of the jet, which contains both substance condensed into clusters of various sizes and a residual gaseous component. By changing the jet flow regimes (parameters of the gas at the nozzle inlet), it is possible to change the weighted average size of the generated clusters, which, in many aspects, determines their properties (including the structural ones). Yet, in the general case, a change in the flow regime will also cause a change in the level of the substance clustering in the jet (the total amount of the clustered substance), which, in turn, will inevitably impact the results of the measurements. In the case of cathodoluminescence studies, this change will affect mostly integrated intensity of the spectral bands, thus making it more difficult to analyze the spectra from the viewpoint of evolution of the spectral characteristics accompanying a change in the average cluster size and in the size-related cluster structure and physical properties. To solve this problem, we propose a new approach to studying luminescence spectra of van-der-Waals clusters, whose basic principles were reported in our short communication [13] based on the experimental results obtained for clusters of argon. Here, we describe the new approach in more detail and communicate more experimental data for clusters of three rare gases – argon, krypton, and xenon.

## 2. Experiment

The paper studies substrate-free nanoclusters of argon, krypton, and xenon produced in gaseous jets adiabatically expanding a into a vacuum through a conical supersonic nozzle with a diameter of 0.34 mm, opening angle of 8.6°, and a ratio of the outlet cross-section to the critical one of 36.7 (a more detailed description of the experimental setup is available in Ref. [14]). By varying the temperature $T_0$ of the gas at the nozzle inlet while keeping the stagnation pressure constant at $p_0$ = 1 atm, we produced beams with clusters of different average sizes and structures. The clusters were excited by electrons in a narrow section of the jet at a distance of 30 mm away from the nozzle outlet, where their average size (average number of atoms per cluster, $\overline{N}$) was already close to the maximum achievable value in a jet for given $p_0$ and $T_0$ and their structure was a thermodynamically equilibrium one. In this section of the jet, the average cluster size $\overline{N}$ can be quantitatively estimated with an error of approximately 30% by using this empirical relation (see, e.g., [15]):

$$\overline{N} = \frac{2\pi}{3}\left(\frac{2R}{a_0}\right)^3 = \gamma\left(\frac{\Gamma^*}{1000}\right)^\chi, \qquad (1)$$

where $\Gamma^*$ is the Hagena parameter, which can be calculated by:

$$\Gamma^* = k_g \left(\frac{0{,}74d}{\tan\alpha}\right)^{0{,}85} \frac{p_0}{T_0^{2{,}29}} = k_g d_{eq}^{0{,}85} \frac{P_0}{T_0^{2{,}29}}. \qquad (2)$$

In these relations, $a_0$ is the cluster crystal lattice parameter, $R$ is cluster radius, $p_0$ (mbar) and $T_0$ (K) are the pressure and temperature of the gas at the nozzle inlet, $d$ (µm) is the critical diameter of the conical nozzle ($d$ = 340 µm), $2\alpha$ is the total cone opening angle ($2\alpha$ = 8.6°), $d_{eq}$ (µm) is the equivalent diameter of the nozzle, and $k_g$ are the constants characteristic of the gases used ($k_{Xe}$= 5500, $k_{Kr}$ = 2890, $k_{Ar}$ = 1650 [16]). The parameters $\gamma$ and $\chi$ in Eq. (1) are different for different supersonic nozzles, we used the values $\gamma$ = 19.5 and $\chi$ = 1.8 that were experimentally verified in the electron diffraction measurements [17] for a nozzle identical to ours in the range $\Gamma^*$ = $10^4$-$10^5$, that is for $\overline{N}$ = 1·10³-8·10⁴ atoms per cluster (at/cl).

The size range of the clusters we studied here was 500-8900 at/cl (cluster diameters being 3.3-8.7 nm) in the case of argon, 100-8500 at/cl (2.1-9.1 nm) in the case of krypton, and 160-18000 at/cl (2.6-12.7 nm) in the case of xenon, all covering both structures with a 5-fold axis of symmetry (amorphous polyicosahedral structure and quasicrystalline structure of the multilayer icosahedron [18, 19]) and the crystalline fcc structure (with an eventual admixture of hcp phase in the largest clusters [20]). The temperature of clusters was virtually constant for all cluster sizes: about 40 K for argon, 60 K for krypton, and 80 K for xenon.

To obtain the emission spectra, we excited clusters with an electron beam of constant-density ($I_{el} \approx 20$ mA) with a subthreshold electron energy $E_{el} = 1$ keV, as well as with photons with energy not exceeding $E_{ph} = 10$ eV. The signal was registered in the VUV spectral range with the photon energies varying 6 to 11.7 eV. This section of the spectrum contains the main molecular continua related to the neutral and charged excited molecular centres (excimer complexes) in Ar, Kr, and Xe clusters.

### 3. Results and discussion

One of the main quantitative parameters of cathodoluminescence spectra is integrated intensity of the spectral features they contain. It provides valuable information on the processes of formation and relaxation of electronic excitations, which, in the case of clusters, largely depend on cluster composition and structure. Since integrated intensity characterizes the total amount of the excited substance clustered in the jet, its value is dependent not only on the physical parameters of one cluster, but also on the total number of clusters in the jet. The cluster size, which also determines the cluster structure, depends on the pressure and temperature of the gas mixture at the nozzle inlet. Changing these parameters, one not only changes the average cluster size in the jet, but also their number. Therefore, to analyze the processes occurring inside a cluster, the measured values of integrated intensity must be normalized, with the normalization factor being the number of clusters in the jet.

Up until now, the contribution of the variable level of substance clustering in the jet to the intensity of spectral features was not taken into account. In this paper, we study the influence of cluster size on cluster spectra and take into consideration the changes in the absolute and relative amount of the clustered substance, which accompany any change in the average cluster size in the jet. Our studies are focused on emission from vibrationally relaxed states of the neutral $(Rg_2)^*$ and charged $(Rg_4^+)^*$ excimer complexes in clusters of three rare gases: Rg = Ar, Kr и Xe. The choice of the bands was largely due to their high intensity in the spectra, which minimizes the possible error in the analysis.

Integrated intensity of an emission band is a function of the excitation cross-section of a unit volume of the emitting substance, the flux density of the incident particles used to excite it, and the probability of implementation of the radiative relaxation channel which is responsible for the formation of the studied band, as well as of the total volume of the substance in which the processes of excitation and subsequent radiation take place. When clusters are excited by a constant-density electron beam, the integrated intensity is proportional to the cluster excitation cross-section, the number of excited clusters, and the probability of the electronic relaxation giving rise to the emission of the analyzed band:

$$I \sim \sigma_{exc} n_{cl} P_{rad}, \tag{3}$$

here $I$ is the integrated intensity of the emission band, $\sigma_{exc}$ is the excitation cross-section of one cluster, $n_{cl}$ is the number of excited clusters, and $P_{rad}$ is the probability of the radiative relaxation channel. The parameter $P_{rad}$ is one that contains the main information about the physics of the relaxation processes which occur in the cluster. To get that information, we also need to know the number of excited clusters and the cluster excitation cross-section.

The number of clusters $n_{cl}$ excited in the jet is determined by the total amount $N_{cl}$ of the clustered substance in the jet and the average cluster size $\overline{N}$ in that section of the jet where it is bombarded by electrons:

$$n_{cl} = N_{cl} / \overline{N}. \tag{4}$$

The parameter $N_{cl}$ is a function of the clustering level of the jet substance:

$$N_{cl} = n_0 \, c_{cl}, \tag{5}$$

$n_0$ being the total number of atoms of the jet substance and $c_{cl}$ being the fraction of atoms condensed into clusters. Then we have:

$$n_{cl} = c_{cl} \, n_0 / \overline{N}. \tag{6}$$

To find $c_{cl}$, we use the results obtained by B.M. Smirnov in Ref. [21], where he estimated theoretically the maximum concentration $c_{max}$ of bound atoms at the end of expansion of pure atomic gas in the jet:

$$c_{max} \sim \frac{T_*}{\varepsilon_0} \ln N_{max}, \tag{7}$$

here $N_{max}$ is the maximum number of atoms in a cluster at the end of the expansion process, $T_*$ is the condensation onset temperature, and $\varepsilon_0$ is the mean binding energy of cluster atoms calculated per one atom. For those clusters that are quite large, the mean binding energy is virtually independent of the cluster size and is close to its value for a macroscopic system, therefore the ratio $\frac{T_*}{\varepsilon_0}$ is pretty much constant.

In our experiments, clusters were excited at a distance from the nozzle outlet where they already had an average size close to the maximum achievable one and their growth was already extremely slow. Thus we can assume the maximum cluster size $N_{max}$ in the jet to be proportional to the weighted average size $\overline{N}$ of the clusters under study, $N_{max} = A\overline{N}$, so the fraction $c_{cl}$ of the clustered substance can be found from Eq. (7) with an accuracy set by the numerical factor $A$ in the logarithmic term, $c_{cl} \sim \ln A\overline{N}$. In this case, Eq. (6) can be presented as follows:

$$n_{cl} \sim n_0 \ln \overline{N} / \overline{N} + n_0 \ln A / \overline{N}, \tag{8}$$

or

$$n_{cl} \sim n_0 \ln \overline{N} / \overline{N} \tag{9}$$

if we take into account that the second term in Eq. (8) is negligibly small for the cluster sizes used in this study.

The value $n_0$ is determined by the stagnation pressure $p_0$ ($n_0 \sim p_0$):

$$n_{cl} \sim p_0 \ln \overline{N} / \overline{N}. \tag{10}$$

In our experiments, the pressure $p_0$ was constant, therefore

$$n_{cl} \sim \ln \overline{N} / \overline{N}. \tag{11}$$

The cross-section of cluster excitation by electrons, in the general case, is a function of electron energy and cluster size. If clusters are excited by electrons having the same energy, the cross-section $\sigma_{exc}$ can be considered a geometrical cross-section of cluster, which is proportional to $\overline{N}^{2/3}$. So, taking into account the dependence $\sigma_{exc} \sim \overline{N}^{2/3}$ and Eq. (11) for $n_{cl}$, we arrive to the following expression for Eq. (3):

$$I \sim \overline{N}^{2/3} n_{cl} P_{rad} \sim \overline{N}^{-1/3} \ln \overline{N}\, P_{rad}. \tag{12}$$

Equation (12) allows us to use the experimentally measured integrated intensity $I$ of a cluster luminescence band to calculate the value of $P_{rad}$, which reflects the physical processes taking place inside a cluster, and thus being behind the radiative channels under study.

To analyze the spectra, it would be convenient to separate the logarithmic and the power-law dependences and use the normalized integrated intensity $I / \ln \overline{N}$:

$$I / \ln \overline{N} \sim \overline{N}^{-1/3} P_{rad}; \qquad P_{rad} \sim \overline{N}^{1/3} I / \ln \overline{N}. \qquad (13)$$

Figure 1 shows examples of cathodoluminescence spectra from crystalline clusters of argon, krypton, and xenon of approximately the same size which contain all spectral features studied in the paper. The most pronounced bands are related to the radiation of the neutral excimer molecules $(Rg_2)^*$ and charged excimer complexes $(Rg_4^+)^*$ inside the clusters (Rg = Ar, Kr, Xe).

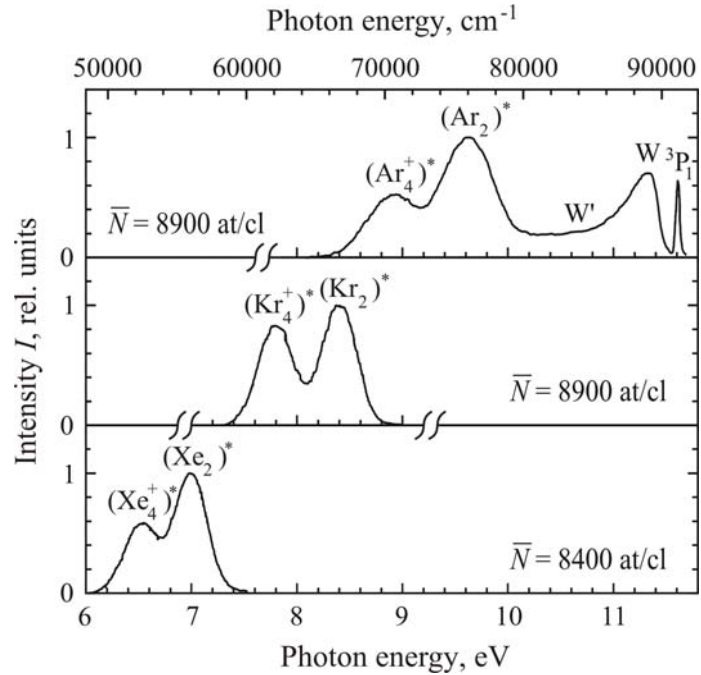

**Fig. 1.** Cathodoluminescence spectra from crystalline clusters of argon, krypton, and xenon with emission bands of the neutral excimers $(Rg_2)^*$ and charged excimer complexes $(Rg_4^+)^*$, Rg = Ar, Kr, Xe.

The $(Rg_2)^*$ bands reflect the transitions from the lowest vibrationally relaxed level ($v = 0$) of the states $^3P_1 + ^1S_0$ ($A\,^3\Sigma_u^+$, $B\,^1\Sigma_u^+$) of the excimer molecule $(Rg_2)^*$ (see, e.g., [14, 22, 23]). This band is formed as a result of the realization of one of the relaxation channels available for excited electrons in rare-gas cryocrystals, both bulk samples and clusters. Impurities and various defects and imperfections of the crystal lattice, including the crystal surface, can greatly intensify the relaxation process. In pure bulk rare-gas solids with a perfect crystalline structure, it is possible to observe at low temperature transitions from both coherent and localized excitonic states. In the case of clusters, however, luminescence at the wavelengths of free excitons was only observed for xenon with a small admixture of argon and explained by formation of exciton-impurity complexes [24]. In luminescence spectra from clusters of pure argon, krypton, and xenon, only emission from localized states has been registered so far, mostly in the form of radiation of the neutral and charged excimer complexes $(Rg_2)^*$ and $(Rg_4^+)^*$. This emission occurs from both crystalline clusters and clusters with an icosahedral structure with no translational symmetry in the radial direction.

In the case of argon clusters, the radiation from the lowest vibrationally relaxed states of the $(Ar_2)^*$ excimer is accompanied by "hot" luminescence related to the radiative transitions from vibrationally excited states. This short-wave contribution can be presented as a superposition of two bands, W′ and W, arising from the transitions from vibrationally partially relaxed and non-relaxed states of the excimer, respectively [14, 25, 26]. They have to be taken into account to allow for more accurate quantitative parameters of the molecular continua $(Ar_2)^*$ and $(Ar_4^+)^*$. The short-wave section of the spectrum ends with a narrow band corresponding to the transition from the excited state $^3P_1$ to the ground state $^1S_0$ in single atoms of argon desorbed from the cluster after excitation by electrons [14], as well as in argon atoms from the gaseous fraction of the jet not condensed into clusters.

The charged complexes $(Rg_4^+)^*$, which are responsible for the formation of intense bands in the long-wave section of the spectrum, result from the interaction between a molecular ion $(Rg_2)^+$ and a neutral excimer molecule $(Rg_2)^*$ [27]. Participation of the ion in the formation of the band is confirmed by the experiments with clusters excited by particles with energies above and below the cluster ionization energy. Figure 2 shows luminescence spectra from clusters of xenon and krypton excited by electrons with an energy of 1 keV and photons with energies not exceeding 10 eV. For both krypton (Fig. 2*a*) and xenon (Fig. 2*b*) clusters, the "red" satellites of the emission bands of the excimer molecules $(Rg_2)^*$ are centred at 7.7 eV and 6.5 eV, respectively, and disappear with the maximum excitation energy dropping below the cluster ionization energy (approx. 12 eV for krypton clusters with sizes about $1·10^3$ at/cl and approx. 10 V for xenon clusters with sizes about $3·10^3$ at/cl [29]).

It is interesting that in bulk samples of argon, krypton, and xenon, even if excited with photons and electrons with energies exceeding that of ionization, no emission from the $(Rg_4^+)^*$ complexes is observed (see, e.g., [30]). This may be due to the fact that when the atom excitation occurs in the inner part of the sample away from its surface, there is a high probability of the atom recombining with the electron before radiating from the $(Rg_4^+)^*$ state. This radiation, however, should be significantly facilitated in the subsurface layers of the sample, because in this case it would be much easier for the electron released during the ionization to leave the sample. Unlike clusters, bulk samples have a relatively small fraction of the subsurface volume and thus the intensity of the $(Rg_4^+)^*$ bands must be much lower than that of the neutral band $(Rg_2)^*$.

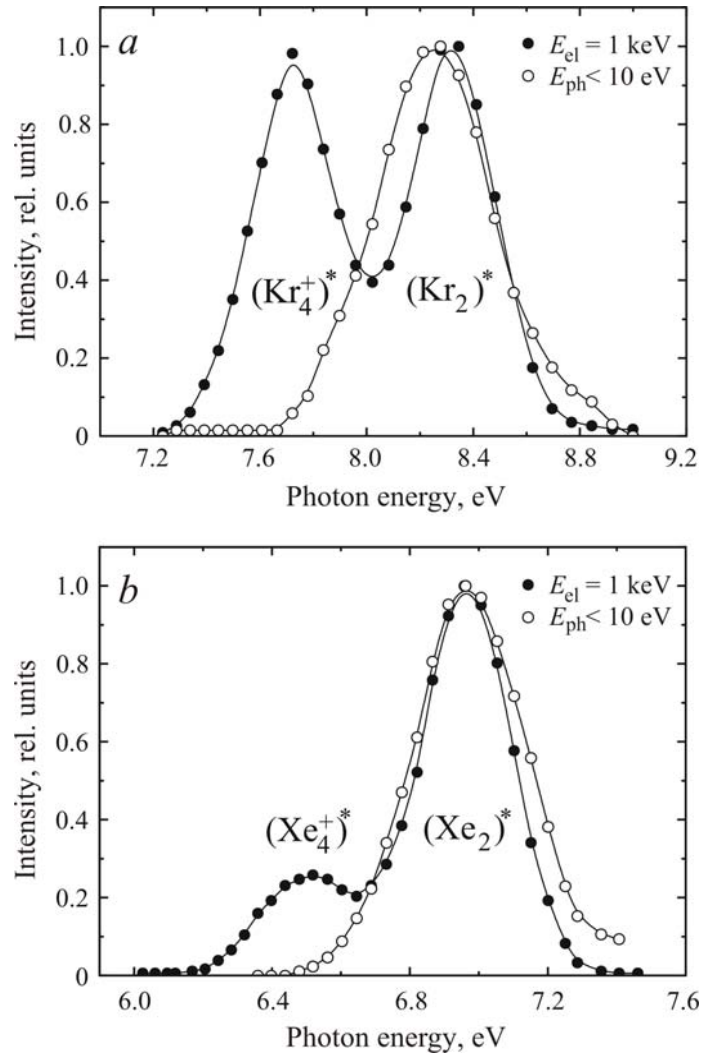

**Fig. 2.** Cathodoluminescence spectra from krypton clusters sized 1100 at/cl (*a*) and xenon clusters sized 3400 at/cl (*b*) in the spectral range of the molecular continua $(Rg_2)^*$ and $(Rg_4^+)^*$ (Rg = Kr, Xe) induced by 1-keV electrons and photons with energies not exceeding 10 eV. A decrease in the energy of incident particles below the cluster ionization energy results in extinction of the emission band of the charged excimer complexes $(Rg_4^+)^*$, which verifies their ionic nature. The data for krypton clusters were taken from Ref. [28], the spectra for xenon clusters were measured in the present study.

Excitation of crystalline clusters of krypton and xenon by photons with energies below 10 eV results in the emission of only one band in the analyzed spectral range – that of the neutral excimer molecules $(Rg_2)^*$, this is shown schematically in Fig. 3*a*. Excitation by electrons with an energy of 1000 eV leads to the radiation of both the neutral and charged excimers from the cluster (Fig. 3*b*). Bombardment of bulk samples by electrons with the same energy (see, e.g., [31]) gives rise to emission dominated by the neutral excimer molecules (Fig. 3*c*). The intensity of the radiation of the neutral excimer molecule $(Rg_2)^*$ is expected to be function of the whole volume of the sample (in other words, the probability of implementation of this radiative channel will be proportional to the total number of atoms making up the sample: $P_{rad} \sim N$), while that of the charged excimer complex $(Rg_4^+)^*$ should be dependent on the volume of the sample subsurface layers ($P_{rad} \sim N^{2/3}$).

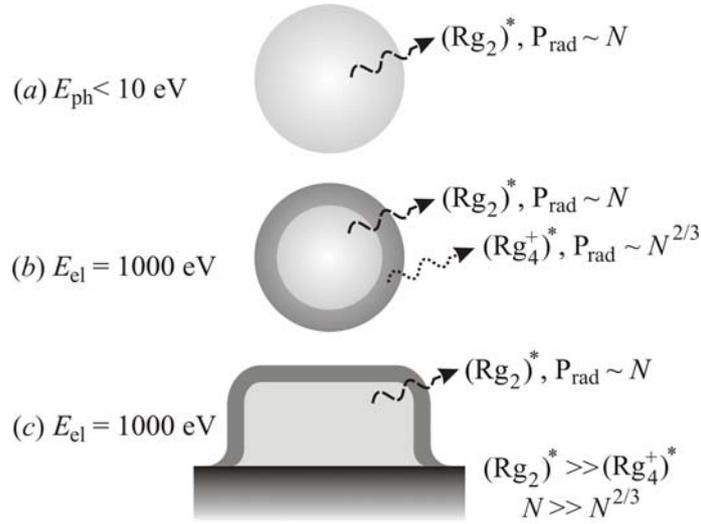

**Fig. 3.** A schematic illustration of radiation of the neutral excimer molecules $(Rg_2)^*$ and charged excimer compelxes $(Rg_4^+)^*$ from the entire volume and some subsurface volume of cluster (*a, b*) and from bulk sample (*c*) of krypton and xenon irradiated by photons with energies $E_{ph} < 10$ eV (*a*) and electrons with an energy $E_{el} = 1$ keV (*b, c*). In bulk samples, only radiation of the $(Rg_2)^*$ molecules is observed because the subsurface volume of the sample is small in comparison with its whole volume.

To verify these relations, we measured cathodoluminescence spectra from substrate-free clusters of argon, krypton, and xenon in a wide range of cluster sizes covering both quasicrystalline structures with a 5-fold axis of symmetry and the crystalline fcc structure.

### 3.1. Argon clusters

Figure 4*a* shows an example of cathodoluminescence spectra from quasicrystalline ($\overline{N} \approx 1000$ at/cl, cluster diameter being 42 Å, upper curve) and crystalline ($\overline{N} \approx 8900$ at/cl, cluster diameter being 87 Å, lower curve) clusters of argon. It also shows decomposition of the spectra into Gaussian components which correspond to the emission band of the vibrationally relaxed neutral excimer molecule $(Ar_2)^*$ at 9.6 eV and that of the charged excimer complex $(Ar_4^+)^*$ at 8.9 eV (dashed lines), as well as to the bands W′ and W with maxima at 10.6 and 11.3 eV, respectively, reflecting the radiation from "hot" vibrational states of the $(Ar_2)^*$ molecule (dotted lines). The "hot" luminescence bands have to be taken into account to allow a more accurate calculation of the integrated intensities of the $(Ar_2)^*$ and $(Ar_4^+)^*$ bands.

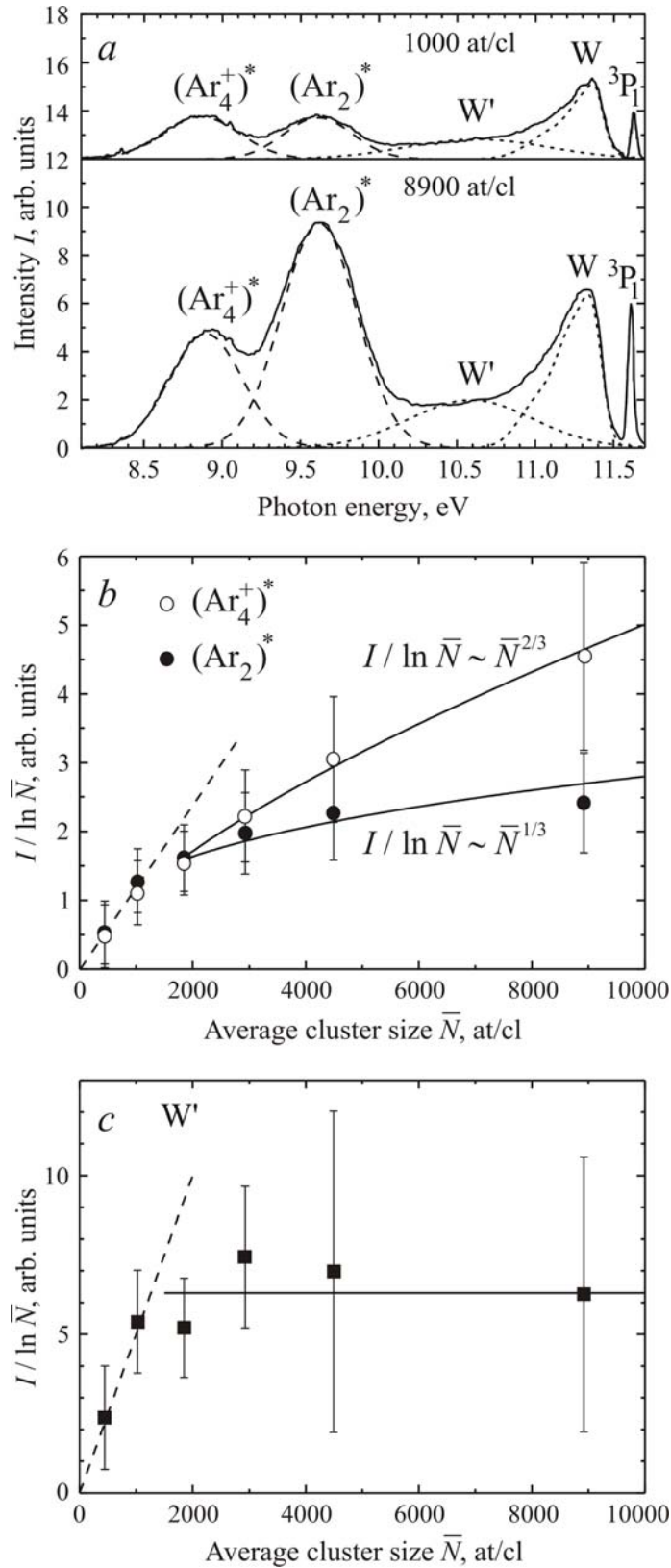

**Fig. 4.** Cathodoluminescence spectra from argon clusters sized $\bar{N} \approx 1000$ and 8900 at/cl (experimental data are solid curves; decomposition of the spectra into Gaussian components is shown as dashed, $(Ar_2)^*$ and $(Ar_4^+)^*$, and dotted, W′ and W, curves; the asymmetric W band is a sum of three Gaussian functions) (*a*). Dependences of the average cluster size $\bar{N}$ of the integrated intensity *I*, normalized by $\ln \bar{N}$, for the bands $(Ar_2)^*$ and $(Ar_4^+)^*$ (*b*) and W′ (*c*): icosahedral clusters with $\bar{N} \leq 1000$ at/cl and fcc clusters with $\bar{N} \geq 1800$ at/cl display different behaviour of the normalized intensity $I / \ln \bar{N}$ (the fitting curves are shown as dashed lines for the icosahedral phase and as solid lines for the fcc phase).

Spectra of bulk samples of argon (see, e.g., [32] for photoexcitation and [33] for electron excitation) do not contain the $(Ar_4^+)^*$ band, while the $(Ar_2)^*$ band is shifted to higher frequencies by approximately 0.1 eV, the relative intensity of the "hot" luminescence (bands W′ and W) is an order of magnitude lower than that in clusters. In the case of bulk samples, the vibrationally relaxed band of the neutral excimer $(Ar_2)^*$ is a superposition of two bands related to the radiative decay of (i) self-trapped excitons (the component at 9.78 eV) and (ii) excitons trapped by structural defects and various imperfections of the crystalline lattice, including those induced by electrons (the component at 9.6 eV) [30]. In clusters, the most efficient channel of exciton trapping is provided by numerous defects formed during the cluster growth, therefore the $(Ar_2)^*$ band in Fig. 4*a* has only the long-wave component at 9.6 eV and can be well described by a single Gaussian. The abnormally high intensity of the "hot" luminescence (W′ and W) is apparently due to the great contribution of molecules desorbed from the clusters (see, e.g., [14]). Since the surface, where desorption usually occurs, is much larger in relative terms in clusters than in bulk samples, the intensity of the bands related to the vibrationally excited states in clusters is also higher.

Evolution of the neutral and ionic molecular continua $(Ar_2)^*$ and $(Ar_4^+)^*$ upon variation in size (and structure) of clusters was studied for argon clusters with the average size ranging from 500 to 8900 at/cl. For each value of the average cluster size, several cathodoluminescence spectra were measured and used for calculation of the average values of integrated intensities $I$ of the bands $(Ar_2)^*$ and $(Ar_4^+)^*$. The size dependences of the intensities normalized by $\ln \overline{N}$ in accordance with Eq. (13) are shown in Fig. 4*b*. We can see that both bands are characterized by two ranges of cluster sizes with different behaviour of the normalized integrated intensity $I / \ln \overline{N}$.

In the case of bigger clusters ($\overline{N} \geq 1800$ at/cl), the normalized intensities tend to grow nonlinearly with $\overline{N}$. This growth can be well described by a power-law function, $I / \ln \overline{N} \sim \overline{N}^\alpha$, with $\alpha = 2/3$ for $(Ar_2)^*$ and $\alpha = 1/3$ for $(Ar_4^+)^*$. Equation (13) allows us to proceed from $I$ to $P_{rad}$, that is from intensity to a parameter which reflects the probability of the relevant molecular emission from one cluster. Then, for the neutral excimer molecules $(Ar_2)^*$ we have $P_{rad} \sim \overline{N}$ and for the charged excimer complexes – $P_{rad} \sim \overline{N}^{2/3}$. This actually means that the probability of radiation of the neutral excimer from an argon cluster is proportional to the total number of atoms in the cluster, that is to the entire cluster volume ($\overline{N} \sim \overline{R}^3$, $P_{rad} \sim \overline{R}^3$, $\overline{R}$ being the average cluster radius), while radiation of the charged complex is function of the number of atoms residing in the subsurface layers of the cluster, the subsurface volume being proportional to the cluster surface ($P_{rad} \sim \overline{R}^{2/3}$). This is in good agreement with the assumption that the radiative decay of the charged excimer complexes should occur mostly in near the sample surface, since it is there that the probability of emission of the electrons released during the ionization is the highest, and thus the probability of recombination of these electrons with the positively charged complexes is the lowest (see Fig. 3).

For cluster size range 1000-1800 at/cl, there is a transition from the two power-law dependencies of the normalized intensities of the bands $(Ar_2)^*$ and $(Ar_4^+)^*$ to a common dependence for both bands which, within the experimental error, can be presented as a linear function: $I / \ln \overline{N} \sim \overline{N}$. The electron diffraction measurements performed on argon clusters demonstrate that clusters with $\overline{N} \leq 2000$ at/cl have structures with a 5-fold axis of symmetry (amorphous polyicosahedral structure and quasicrystalline structure of multilayer icosahedron) [19, 20], while larger clusters are characterized by an fcc structure (with a possible admixture of the hcp phase in clusters with the number of atoms of the order of $10^4$) [20]. This suggests that the different behaviour of the normalized integrated intensity of the $(Ar_2)^*$ and $(Ar_4^+)^*$ bands can be due to the difference in the cluster structure. The transitional region ($\overline{N}$ = 1000-1800 at/cl) thus reflects the simultaneous presence in the jet of clusters with a crystalline (fcc) structure and those with a quasicrystalline (icosahedral) structure. The latter, which are characterized by the linear dependence of the normalized integrated intensity of both molecular continua, prevail in smaller clusters ($\overline{N} \leq 1000$ at/cl).

The additional deviation towards lower values of the normalized intensities of the $(Ar_2)^*$ and $(Ar_4^+)^*$ bands for icosahedral clusters ($\overline{N} \leq 1000$ at/cl) with respect to how they behave in the case of fcc phase can be qualitatively explained by a greater binding energy $\varepsilon_0$ of atoms in the cluster with an icosahedral structure [34], since a variable $\varepsilon_0$ in Eq. (7) implies a different expression for Eq. (13): $I / \ln \overline{N} \sim \overline{N}^{-1/3} P_{rad} / \varepsilon_0$. The close values of the intensity of both bands are most probably due to the fact that in small clusters the subsurface layers with a facilitated emission of electrons occupy a significant volume of the cluster and thus it hardly makes any sense to speak in terms of *surface* and *core*: the largest clusters from this size range have 900-1000 atoms per cluster and are characterized by the structure of multilayer icosahedron [18], which means they are made up of only 6 atomic layers (Mackay spheres) around the central atom.

The cluster structure also affects the "hot"-luminescence band W′, which reflects transitions from partially vibrationally excited levels of the excited neutral excimer molecule $(Ar_2)^*$ (see Fig. 4*d*). The contributions to the band may come from both excimer molecules inside the cluster and those desorbed from it (see [14, 25]). The complexity of unequivocal identification of the band profile complicates analysis of the size dependence of its normalized intensity: in the size range of crystalline clusters (1800-8900 at/cl), it can be considered constant (shown in Fig. 4*d*) or, alternatively, described satisfactorily within the experimental error by various power-law functions (or, rather, by a linear combination of several power-law functions, as the band's nature would suggest). Anyway, a qualitative analysis of the W′ band's behaviour allows us to believe that, just like the emission bands of the vibrationally relaxed molecules $(Ar_2)^*$ and $(Ar_4^+)^*$, it is sensitive to the cluster structure and indicates a transition from the quasicrystalline phase to the crystalline one in the same cluster size range – 1000-1800 at/cl.

## 3.2. Krypton clusters

Analysis of the integrated intensities of the emission bands of the neutral and charged excimer complexes in krypton clusters is shown in Figs. 5 and 6. An example of the cathodoluminescence spectra in the frequency range of the spectral bands is shown in Fig. 5*a* for a wide range of cluster sizes, which covers both icosahedral and fcc phases. Similarly to argon clusters, the radiation intensity of the charged complexes is comparable to that of the neutral excimer molecules, this is also the case for large clusters.

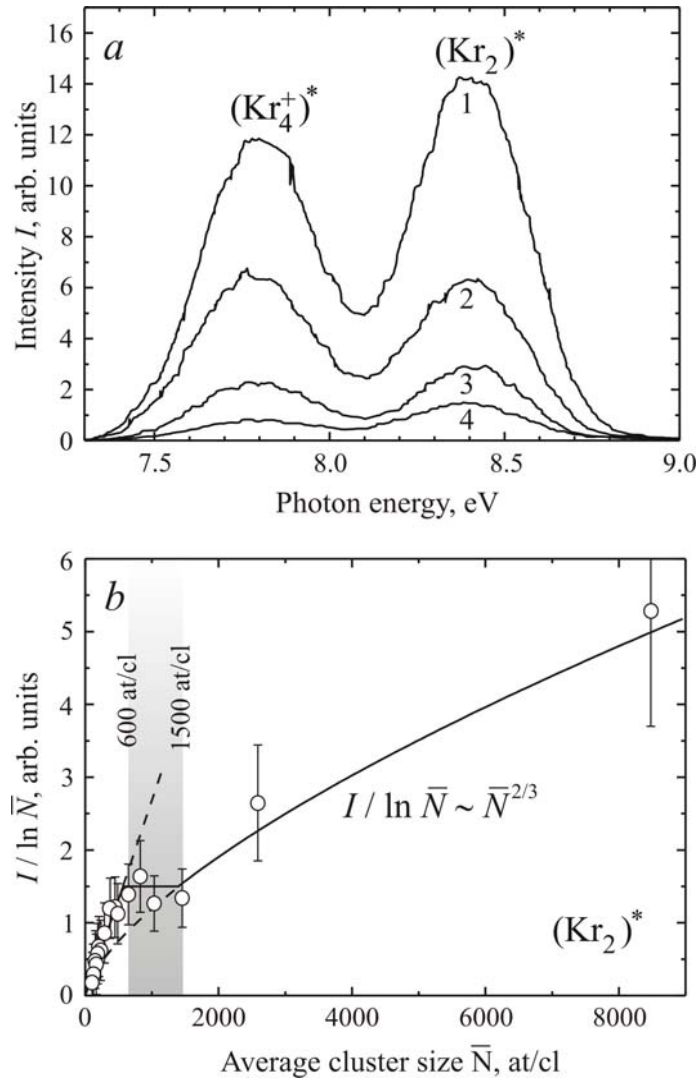

Fig. 5. Cathodoluminescence spectra in the spectral range of the $(Kr_2)^*$ and $(Kr_4^+)^*$ bands for crystalline (fcc) krypton clusters with 8500 atoms per cluster (1) and 2600 atoms per cluster (2) and quasicrystalline (icosahedral) clusters sized 1000 at/cl (3) and 290 at/cl (4) (*a*). Normalized intensity $I/\ln \overline{N}$ of the band $(Kr_2)^*$ versus average cluster size $\overline{N}$ (*b*).

The normalized integrated intensity curve of the $(Kr_2)^*$ band averaged over a series of measurements is shown in Fig. 5*b* for the entire range of cluster sizes studied (100-8500 at/cl). Just like in the case of argon clusters, here too we can see two ranges of cluster sizes with qualitatively different behaviour of the intensity and an intermediate range in between. For clusters with $\overline{N} \geq 1500$ at/cl, the normalized intensity is proportional to $\overline{N}^{2/3}$, which, see Eq. (13), implies the probability of radiation of the neutral molecule $(Kr_2)^*$ from inside a cluster being proportional to the whole cluster volume ($P_{rad} \sim \overline{N}$). In the case of small clusters with $\overline{N} \leq 600$ at/cl, the normalized intensity can be seen, within the experimental and calculating error, as a linear function. The transitional region between the two types of curves thus refers to clusters with the average size ranging from 600 to 1500 at/cl. This corresponds with the electron diffraction data [35], according to which clusters of krypton sized some 2000 at/cl and more have predominantly fcc structure, while smaller clusters have a broad intermediate size range between the quasicrystalline phase with an icosahedral structure and the crystalline fcc phase.

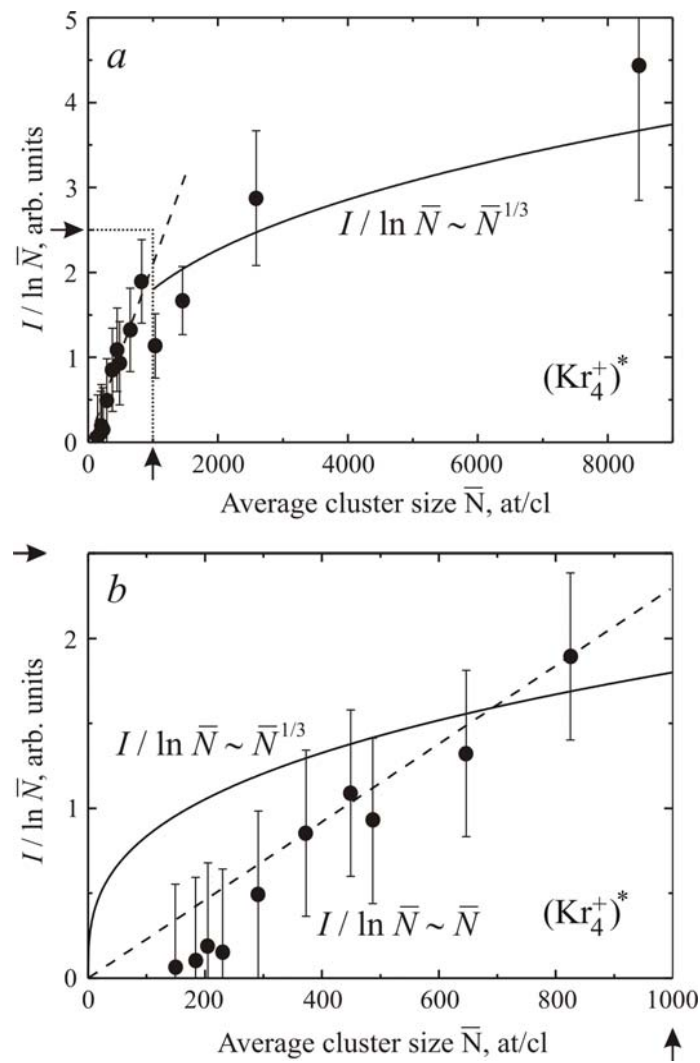

Fig. 6. Normalized intensity $I / \ln \overline{N}$ of the band $(Kr_4^+)^*$ versus average cluster size in the whole range of cluster size (*a*) and for small clusters (*b*) in the cluster size range marked by dotted lines and arrows in (*a*).

The data on the normalized integrated intensity of the emission band of the charged excimer complexes $(Kr_4^+)^*$ for the whole range of cluster sizes are reported in Fig. 6. Here, for clusters with $\overline{N} \geq 1500$ at/cl, the normalized intensity can be described by the function $I/\ln\overline{N} \sim \overline{N}^{1/3}$, which means that the probability of radiation of the $(Kr_4^+)^*$ band, in analogy with the $(Ar_4^+)^*$ band, is proportional to the cluster surface area and, therefore, the charged excimer complexes in krypton radiate most effectively from some subsurface volume of the crystalline cluster.

If we take a look at small clusters, $\overline{N} \leq 600$ at/cl, we will see that the normalized intensity changes its behaviour, like it is shown on a larger scale in Fig. 6b. In virtually all this region, corresponding to the icosahedral cluster structure, the normalized intensity data fall into a linear fit line: $I/\ln\overline{N} \sim \overline{N}$.

### 3.3. Xenon clusters

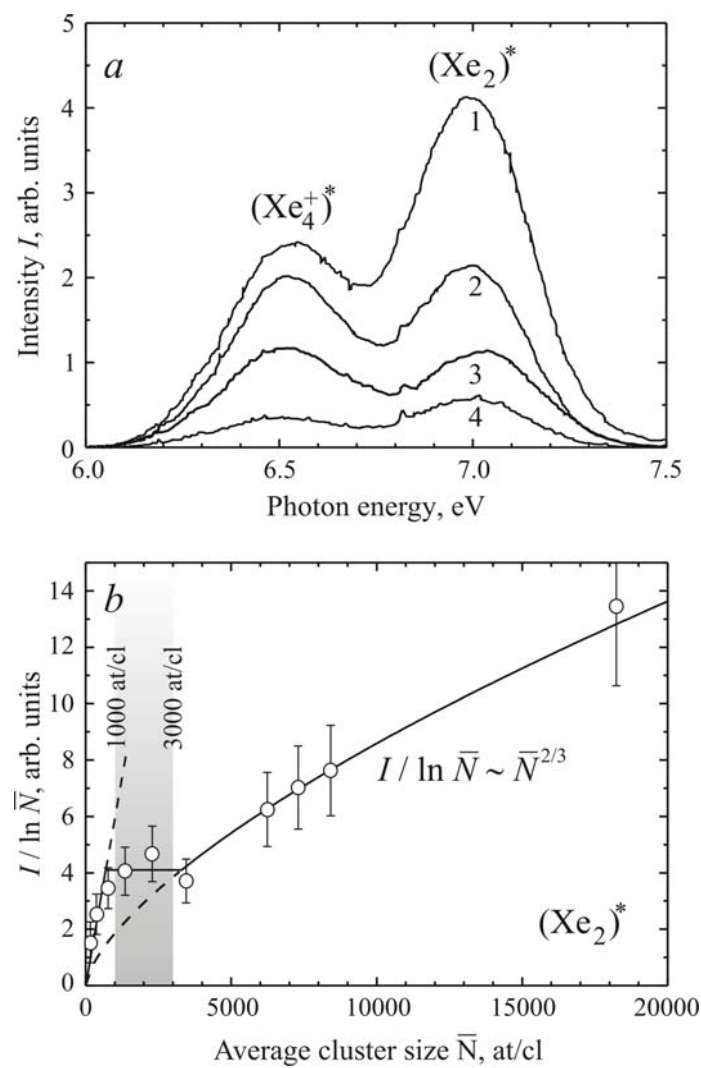

Fig. 7. Cathodoluminescence spectra in the spectral range of the $(Xe_2)^*$ and $(Xe_4^+)^*$ bands for crystalline (fcc) xenon clusters with 18000 atoms per cluster (1), 6200 atoms per cluster (2), and 3500 atoms per cluster (3) and quasicrystalline (icosahedral) clusters sized 370 at/cl (4) (a). Normalized intensity $I/\ln\overline{N}$ of the band $(Xe_2)^*$ versus average cluster size $\overline{N}$ (b).

Let us now consider the similar measurements for clusters of xenon (see Figs. 7 and 8). Typical cathodoluminescence spectra with the molecular continua of the neutral and charged $(Xe_2)^*$ and $(Xe_4^+)^*$ for icosahedral and crystalline clusters of different sizes are shown in Fig. 7a. The emission band of the charged excimer complexes also has a significant intensity which is close to that of the band of the neutral excimers. The curves of the normalized intensity of the $(Xe_4^+)^*$ band are shown in Fig. 7b for clusters with sizes ranging from 160 to 18000 at/cl. We can see three cluster size ranges, each with its own intensity evolution: (i) a size range of crystalline clusters ($\overline{N} \geq 3000$ at/cl) with $I / \ln \overline{N} \sim \overline{N}^{2/3}$, which implies that this type of radiation is formed in the entire cluster volume; (ii) a size range of icosahedral clusters ($\overline{N} \leq 1000$ at/cl) with a linear intensity, which reflects a change in the atom binding energy caused by the change in cluster structure, and (iii) an intermediate size range ($1000 \leq \overline{N} \leq 3000$ at/cl), which suggests that clusters of both phases are present in the jet.

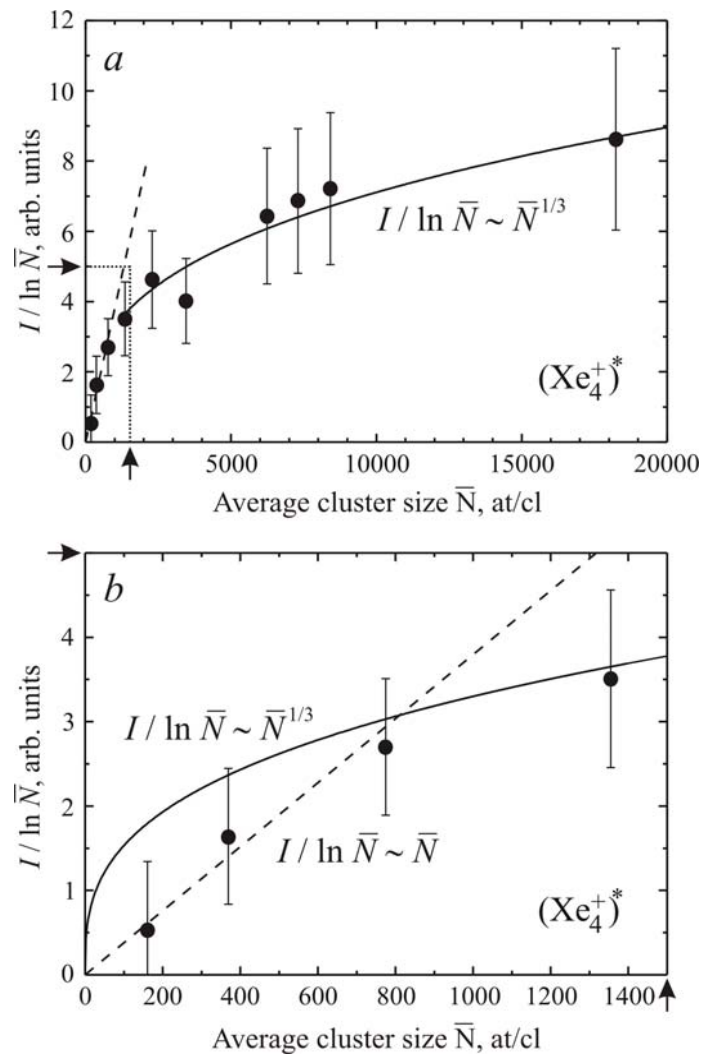

Fig. 8. Normalized intensity $I / \ln \overline{N}$ of the band $(Xe_4^+)^*$ versus average cluster size in the whole range of cluster size (*a*) and for small clusters (*b*) in the cluster size range marked by dotted lines and arrows in (*a*).

The evolution of the $(Xe_4^+)^*$ band is shown in Fig. 8. The function $I / \ln \overline{N} \sim \overline{N}^{1/3}$, typical of the molecules emitting from subsurface layers is observed for crystalline clusters (Fig. 8*a*), while the linear behaviour of the normalized intensity of the band related to the charged excimer complexes is thus characteristic of icosahedral clusters (Fig. 8*b*). For both krypton and xenon clusters, the experimental data did not allow us to clearly establish the cluster size range which is "transitional" between the two structures. Yet, the transition itself is quite evident and falls into the average size range that follows from the analysis of the bands of the neutral excimers in these clusters.

## 4. Conclusions

The present study deals with cathodo- and photoluminescence spectra in the VUV frequency range of the emission bands of the neutral and charged excimer complexes $(Rg_2)^*$ and $(Rg_4^+)^*$ in substrate-free nanoclusters of argon, krypton, and xenon with average sizes varying from 100 to 18000 atoms per cluster to include the cluster size ranges of both quasicrystalline icosahedral and crystalline fcc structures. To quantitatively analyze the integrated intensity *I* of these bands, we apply our new approach which takes into account the amount of the clustered substance in the supersonic jet.

The obtained experimental results allow us to demonstrate that the fraction $c_{cl}$ of the substance condensed into clusters is proportional to logarithm of the average cluster size $\overline{N}$ in the jet, $c_{cl} \sim \ln \overline{N}$. This finding makes it possible to use the cathodoluminescence spectra to establish the cluster size ranges which correspond to quasicrystalline clusters with a 5-fold axis of symmetry (multilayer icosahedron and amorphous polyicosahedral structure) and crystalline fcc clusters. The proposed technique can also be used to study the intermediate cluster size range with both icosahedral and fcc clusters present in the jet.

We find that in crystalline clusters of argon, krypton, and xenon the radiation of the vibrationally relaxed neutral excimer molecules $(Rg_2)^*$ occurs from within the whole cluster ($I / \ln \overline{N} \sim \overline{N}^{2/3}$), while the charged excimer complexes $(Rg_4^+)^*$ emit from some subsurface layers ($I / \ln \overline{N} \sim \overline{N}^{1/3}$).

The obtained results suggest that the proposed approach can be used for spectroscopic analysis of more complex rare-gas systems and some simple molecular substances, which can also shed light on their structure. The new approach is expected to be helpful in opening up new ways for complex studies of nanoclusters which combine spectroscopic (cathodoluminescence) and structural (electron diffraction) techniques.

The work was supported by the National Academy of Sciences of Ukraine (grant 6/15-H) and NATO SPS Programme (SPS.MYP 985481).